\documentclass[12pt]{article}
\usepackage{amssymb,amsmath,epsfig}
\usepackage{graphicx}
\begin{document}

\title{\bf Complexity Factor for Static Sphere in Self-interacting Brans-Dicke Gravity}

\author{M. Sharif \thanks{msharif.math@pu.edu.pk} and Amal Majid
\thanks{amalmajid89@gmail.com}\\
Department of Mathematics, University of the Punjab,\\
Quaid-e-Azam Campus, Lahore-54590, Pakistan.}

\date{}

\maketitle
\begin{abstract}
In this paper, we study the complexity factor of a static
anisotropic sphere in the context of self-interacting Brans-Dicke
theory. We split the Riemann tensor using Bel's approach to obtain
structure scalars relating to comoving congruence and Tolman mass in
the presence of a scalar field. We then define the complexity factor
with the help of these scalars to demonstrate the complex nature of
the system. We also evaluate the vanishing complexity condition to
obtain solutions for two stellar models. It is concluded that the
complexity of the system increases with the inclusion of the scalar
field and potential function.
\end{abstract}

\section{Introduction}

The study of large scale structures such as stars, galaxies and
their clusters provide insights into the dynamics of the universe.
The intricate nature of these stellar structures massively depends
upon the interdependent physical variables such as energy density,
pressure and heat flux. To determine the characteristics of
celestial objects, it is necessary to gain complete or maximum
information to compute the complexity factor for the considered
system. Such a factor specifies the degree of complexity found
within the system and provides a measure for comparing different
self-gravitating structures. The problem of measuring the complexity
of cosmic structures is not new and numerous attempts have been made
to define such a factor \cite{1}. In spite of these works, a
consensus on a definition of complexity factor has not been
achieved.

Among the proposed definitions, the concepts of order or arrangement
of atoms and entropy of a system have been taken into account.
However, the need for a complexity factor depending on other
variables arises when two simple but different models in physics, a
perfect crystal, and an ideal gas, are compared. The crystal is a
symmetric alignment of ordered atoms whose probability distribution
of accessible states depends on the perfect symmetry. Therefore,
minimal information is required to describe the distances and
symmetries of its elementary cell. On the other hand, the ideal gas
is completely orderless having the same probability for any of its
accessible states leading to maximum information for its
description. This diversity in the structure and pattern of two
simple models motivated Lopez-Ruiz et al. \cite{5} to update the
definition of complexity factor by including the notion of
disequilibrium. The abstract idea was to measure how different
probabilistic states differ from the equiprobable distribution of
the system. This led to a new definition of complexity factor which
yielded zero complexity for, both, the perfect crystal and the ideal
gas.

Based on the definition of Lopez-Ruiz et al., the complexity factor
for the self-gravitating compact structures (such as neutron stars
and white dwarfs) has been computed by using energy density in place
of probability distribution \cite{9}. However, the definition of
complexity discussed above takes only energy density into account
neglecting other variables (pressure, temperature, dissipation,
etc.) that also play a vital role in structure formation. Recently,
Herrera \cite{13} proposed a new technique to compute complexity for
static spherical symmetry in the context of general relativity (GR)
by overcoming this deficiency. This definition is characterized by
the inhomogeneous energy density, active gravitational mass and
anisotropic pressure of the fluid rather than the information and
disequilibrium of the structure. The orthogonal splitting of the
Riemann tensor was employed to obtain a scalar function which
incorporated the effects of the mentioned physical quantities. This
scalar was termed as complexity factor which vanished in the case of
homogeneous energy density and isotropic pressure. Herrera et al.
\cite{13*} extended this definition to dynamical spheres by
including the condition of minimal complexity to discuss the pattern
of evolution. Sharif and Butt \cite{14*} used the same technique to
evaluate the complexity factor for a charged sphere as well as
static cylinder and discussed the vanishing complexity conditions in
both cases. The complexity factor has also been investigated for the
case of axially symmetric static source \cite{15*}.

Recent cosmic observations such as redshift and distance-luminosity
relationship of type IA Supernovae \cite{14} led to the discovery of
the accelerated expansion of the cosmos. The current scenario of the
universe within the framework of GR leads to problems like
fine-tuning and cosmic coincidence. This triggered the modification
of this theory to include the effects of accelerated expansion.
Brans-Dicke (BD) theory, based on Mach's principle \cite{15}, is one
of the scalar-tensor generalizations of the Einstein-Hilbert action
that has provided an effective explanation of the expanding
universe. In this theory, an additional scalar field $\Phi$ is
introduced alongside the metric tensor which contributes to the
geometry of spacetime with no effect on matter. The scalar field is
taken to be the reciprocal of dynamical gravitational constant $G$
and is coupled to matter as well as gravity through a coupling
constant $\omega_{BD}$. Further, this constant can be treated as a
parameter that can be tuned to get the desired results.

The inflationary model given by BD gravity holds for lower values of
$\omega_{BD}$ \cite{17} whereas the local gravity tests of the solar
system are satisfied for the constrained values of
$\omega_{BD}\geq40,000$ \cite{16}. To resolve this conflict, this
theory is modified by including a scalar potential $V(\Phi)$
yielding self-interacting BD theory \cite{18}. This function allows
an effective mass of the field to solve the inconsistency in values
of $\omega_{BD}$ obtained for a weak field at the cosmic scale.
Sharif and Manzoor \cite{22} studied the dynamics of
self-gravitating fluids in BD theory with non-zero potential and
concluded that models for regular distribution of scalar field are
consistent with GR. They also discussed the evolution of a sphere by
incorporating orthogonal splitting of the Riemann tensor \cite{23}.
Diverse geometries have been explored by constructing the structure
scalars in the background of self-interacting BD theory \cite{23'}.

Recently, Abbas and Nazar \cite{23*} analyzed the vanishing
complexity condition in the context of $f(R)$ gravity. In this
paper, we study the complexity factor for a static anisotropic
sphere in self-interacting BD theory. This paper is organized as
follows. In section \textbf{2}, we establish field equations and
physical variables describing the anisotropic matter distribution.
Section \textbf{3} gives an overview of orthogonal splitting of the
Riemann tensor and the resultant structure scalars. The complexity
factor formulated from these scalars is introduced in section
\textbf{4}. We also present a solution to the field equations by
using the zero complexity condition. In the last section, we
summarize our results.

\section{Self-interacting Brans-Dicke Theory and Matter Variables}

The action of self-interacting BD theory \cite{18} with $8\pi G_0=1$
is defined as
\begin{equation}\label{0}
S=\int\sqrt{-g}(R\Phi-\frac{\omega_{BD}}{\Phi}\nabla^{\mu}\nabla_{\mu}\Phi
-V(\Phi)+\emph{L}_m)d^{4}x,
\end{equation}
where $g,~R$ and $\emph{L}_m$ represent the determinant of the
metric tensor, Ricci scalar and matter Lagrangian, respectively. The
second term in Eq.(\ref{0}) represents the kinetic term which
indicates the presence of a singularity due to $\frac{1}{\Phi}$.
However, the term can be expressed in canonical form by introducing
a new field $\varphi$ and a constant $\varpi=\pm1$ \cite{18**}. The
positive value of $\varpi$ corresponds to a normal field with
positive energy, i.e., it is not a ghost field. The variation of the
above action with respect to $g_{\mu\nu}$ and $\Phi$ yields the BD
field equations and evolution equation, respectively, given as
\begin{eqnarray}\label{1}
G_{\mu\nu}&=&T^{\text{(eff)}}_{\mu\nu}=\frac{1}{\Phi}(T_{\mu\nu}^{(m)}
+T_{\mu\nu}^\Phi),\\\label{2}
\Box\Phi&=&\frac{T^{(m)}}{3+2\omega_{BD}}+\frac{1}{3+2\omega_{BD}}
(\Phi\frac{dV(\Phi)}{d\Phi}-2V(\Phi)),
\end{eqnarray}
where the energy-momentum tensor $T_{\mu\nu}^{(m)}$ represents the
matter distribution and $T^{(m)}$ is its trace. The energy-momentum
tensor of scalar field has the following form
\begin{equation}\label{3}
T_{\mu\nu}^\Phi=\Phi_{,\mu;\nu}-g_{\mu\nu}\Box\Phi+\frac{\omega_{BD}}{\Phi}
(\Phi_{,\mu}\Phi_{,\nu}
-\frac{g_{\mu\nu}\Phi_{,\alpha}\Phi^{,\alpha}}{2})-\frac{V(\Phi)g_{\mu\nu}}{2},
\end{equation}
where
$\Box\Phi=\Phi^{,\mu}_{~;\mu}=(-g)^{-\frac{1}{2}}[(-g)^{\frac{1}{2}}
\Phi^{,\mu}]_{,\mu}$ with $\Box$ being the d'Alembertian operator.

We consider a static sphere bounded by a hypersurface $\Sigma$
defined by
\begin{equation}\label{4}
ds^2=e^{\nu(r)}dt^2-e^{\lambda(r)}dr^2-r^2(d\theta^2+\sin^2\theta
d\phi^2).
\end{equation}
The interior of the sphere is filled with anisotropic fluid
described by
\begin{equation}\label{5}
T_{\nu}^{\mu(m)}=\rho u^{\mu}u_{\nu}-Ph^{\mu}_{\nu}+\Pi_{\nu}^{\mu},
\end{equation}
where $u^\mu$ is the 4-velocity of comoving observer and $\rho$
denotes the energy density. The anisotropy in pressure is
represented by $P$ and $\Pi_{\nu}^{\mu}$ which satisfy the following
relations
\begin{eqnarray*}
\Pi_{\nu}^{\mu}&=&\Pi(s^{\mu}s_{\nu}+\frac{h^{\mu}_{\nu}}{3}),
\quad P=\frac{1}{3}(P_{r}+2P_{\perp}),\\
\Pi&=&P_r-P_\perp,
\quad h^{\mu}_{\nu}=\delta^{\mu}_{\nu}-u^{\mu}u_{\nu},\\
u^\mu&=&(e^{\frac{-\nu}{2}},0,0,0),\quad
s^\mu=(0,e^{\frac{-\lambda}{2}},0,0),
\end{eqnarray*}
where $s^{\mu}$ is a unit 4-vector in the radial direction which
satisfies the following conditions
\begin{eqnarray*} s^\mu u_\mu=0,\quad s^\mu
s_\mu=1.
\end{eqnarray*}
Here $P_r$ and $P_\perp$ denote radial and tangential pressure,
respectively. Using Eqs.(\ref{1}) and (\ref{3})-(\ref{5}), the field
equations are obtained as
\begin{eqnarray}\nonumber
\frac{1}{r^2}-e^{-\lambda}\left(\frac{1}{r^2}-\frac{\lambda'}{r}\right)&=&
\frac{1}{\Phi}\left\{\rho+e^{-\lambda}\left[\left(-\frac{2}{r}-\frac{\lambda'}{2}
\right)\Phi'\right.\right.\\\label{6}
&+&\left.\left.\Phi''+\frac{\omega_{BD}}{2\Phi}\Phi'^2-e^\lambda\frac{V(\Phi)}
{2}\right]\right\},\\\nonumber
-\frac{1}{r^2}+e^{-\lambda}\left(\frac{1}{r^2}+\frac{\nu'}{r}\right)&=&
\frac{1}{\Phi}\left\{P_r+e^{-\lambda}\left[\left(-\frac{2}{r}-\frac{\nu'}
{2}\right)\Phi'\right.\right.\\\label{7}
&+&\frac{\omega_{BD}}{2\Phi}\Phi'^2+\left.\left.e^\lambda\frac{V(\Phi)}{2})\right]
\right\},\\\nonumber
\frac{e^{-\lambda}}{4}\left(2\nu''+\nu'^2-\lambda'\nu'+2\frac{\nu'-\lambda'}{r}\right)&=&
\frac{1}{\Phi}\left\{P_\perp+e^{-\lambda}\left[-\left(\frac{1}{r}-\frac{\lambda'}
{2}+\frac{\nu'} {2}\right)\Phi'\right.\right.\\\label{8}
&-&\left.\left.\Phi''-\frac{\omega_{BD}}{2\Phi}\Phi'^2+e^\lambda\frac{V(\Phi)}{2}
\right]\right\},
\end{eqnarray}
and the wave equation (\ref{2}) takes the form
\begin{eqnarray}\nonumber
\Box\Phi&=&-e^{-\lambda}\left[\left(\frac{2}{r}-\frac{\lambda'} {2}
+\frac{\nu'}{2}\right)\Phi'+\Phi''\right]\\\label{2*}
&=&\frac{1}{3+2\omega_{BD}}\left[\rho-3P+
\left(\Phi\frac{dV(\Phi)}{d\Phi}-2V(\Phi)\right)\right],
\end{eqnarray}
prime denotes differentiation with respect to $r$.

The spacetime is divided by the hypersurface $\Sigma$ into two
different regions, interior and exterior. The exterior region is
taken to be the Schwarzschild spacetime given by
\begin{equation*}
ds^2=(1-\frac{2M}{r})dt^2-\frac{1}{(1-\frac{2M}{r})}dr^2
-r^2(d\theta^2+\sin^2\theta d\phi^2),
\end{equation*}
where $M$ is the mass of the exterior region. To ensure smoothness
and continuity of geometry at the boundary surface
($r=r_{\Sigma}=\text{constant}$), the following conditions must be
satisfied
\begin{eqnarray*}
(e^{\nu})_{\Sigma}&=&(1-\frac{2M}{r})_{\Sigma},\\
(e^{-\lambda})_{\Sigma}&=&(1-\frac{2M}{r})_{\Sigma},\\
(P_{r})_{\Sigma}&=&0.
\end{eqnarray*}
 The total energy within a sphere of radius $r$ is
computed through the Misner-Sharp formula \cite{18*} which yields
\begin{equation}\label{9}
m=\frac{r}{2}R^3_{232}=\frac{r}{2}(1-e^{-\lambda})=\frac{1}{2}\int
r^{2}T^{0\text{(eff)}}_{0}dr.
\end{equation}
The Tolman-Oppenheimer-Volkoff equation is obtained through the
field equations and mass function as
\begin{equation}\label{8*}
T^{1'(\text{eff})}_{1}=\frac{2m-r^3T^{1(\text{eff})}_{1}}{2r(r-2m)}
(T^{0(\text{eff})}_{0}-T^{1(\text{eff})}_{1})
+\frac{2}{r}(T^{2(\text{eff})}_{2}-T^{1(\text{eff})}_{1}).
\end{equation}
The mass function can be expressed in terms of the Weyl tensor which
evaluates the effect of tidal forces and appears as the traceless
part in the splitting of Riemann tensor as
\begin{equation}\label{1'}
R^{\mu}_{\alpha\beta\sigma}=C^{\mu}_{\alpha\beta\sigma}+\frac{R^{\mu}_{\beta}}
{2}g_{\alpha\sigma}-\frac{R_{\alpha\beta}}{2}\delta^{\mu}_{\sigma}
+\frac{R_{\alpha\sigma}}{2}\delta^{\mu}_{\beta}
-\frac{R^{\mu}_{\sigma}}{2}g_{\alpha\beta}-\frac{1}{6}(\delta^{\mu}_{\beta}
g_{\alpha\sigma}-g_{\alpha\beta}\delta^{\mu}_{\sigma}),
\end{equation}
where $C^{\mu}_{\alpha\beta\sigma}$ and $~R_{\alpha\beta}$ are the
Weyl tensor and Ricci tensor, respectively.

The Weyl tensor is decomposed into trace-free electric
($E_{\alpha\beta}$) and magnetic ($H_{\alpha\beta}$) parts by using
the 4-velocity of the observer. In the case of spherical symmetry,
these tensors reduce to
\begin{eqnarray}\label{11}
E_{\alpha\beta}&=&C_{\alpha\mu\beta\sigma}u^{\mu}u^{\sigma},\\\label{11*}
H_{\alpha\beta}&=&0,
\end{eqnarray}
where
\begin{eqnarray}\nonumber
C_{\mu\nu\kappa\sigma}&=&(g_{\mu\nu\alpha\beta}g_{\kappa\sigma\delta\gamma}-
\eta_{\mu\nu\alpha\beta}\eta_{\kappa\sigma\delta\gamma})u^{\alpha}u^{\delta}
E^{\beta\gamma},\\\label{12}
g_{\mu\nu\alpha\beta}&=&g_{\mu\alpha}g_{\nu\beta}-g_{\mu\beta}g_{\nu\alpha},
\end{eqnarray}
and $\eta_{\mu\nu\alpha\beta}=u_{\mu}\epsilon_{\nu\alpha\beta}$ is
the Levi-Civita tensor where $\epsilon_{\nu\alpha\beta}$ is the
permutation symbol. Substituting the Weyl tensor in Eq.(\ref{11}),
it follows that
\begin{equation}\label{13}
E_{\alpha\beta}=\varepsilon(s_{\alpha}s_{\beta}+\frac{h_{\alpha\beta}}{3}),
\end{equation}
with
\begin{eqnarray}\label{14}
&&\varepsilon=\frac{e^{-\lambda}}{4}\left(-\nu''-\frac{\nu'^2-\lambda'\nu'}{2}+\frac{\nu'-\lambda'}
{r}+2\frac{1-e^{\lambda}}{r^2}\right),\\\nonumber
&&E^{\mu}_{\mu}=0=E_{\mu\nu}u^{\nu}.
\end{eqnarray}
Through Eqs.(\ref{1}) and (\ref{9}), we obtain the relation
\begin{equation}\label{15}
m=\frac{r^3}{6}
(T^{0\text{(eff)}}_{0}-T^{2\text{(eff)}}_{2}+T^{1\text{(eff)}}_{1})+\frac{\varepsilon
r^3}{3},
\end{equation}
leading to a definition for $\varepsilon$ given by
\begin{equation}\label{16}
\varepsilon=-\frac{1}{2r^3}\int_{0}^{r}r^{3}T^{0'\text{(eff)}}_{0}dr+
\frac{1}{2}(T^{2\text{(eff)}}_{2}-T^{1\text{(eff)}}_{1}).
\end{equation}

This demonstrates the relationship between the Weyl tensor,
inhomogeneous energy density and anisotropic pressure in the
presence of scalar field. Substituting the above equation in
(\ref{15}), the mass function can be rewritten as
\begin{equation}\label{17}
m(r)=\frac{r^3}{6}T^{0\text{(eff)}}_{0}-\frac{1}{6}\int_{0}^{r}r^{3}
T^{0'\text{(eff)}}_{0}dr.
\end{equation}
It is observed that the first term on the right side gives the value
of mass function when the energy density is homogeneous whereas the
second term exhibits the change induced by the inhomogeneous energy
density. We now find the total mass of the spherical system enclosed
within the boundary $r_{\Sigma}$ using an alternate definition
proposed by Tolman \cite{24} as
\begin{equation}\label{18}
m_T=\frac{1}{2}\int_{0}^{r}r^{2}e^{\frac{\nu+\lambda}{2}}(T^{0\text{(eff)}}_{0}
-T^{1\text{(eff)}}_{1}-2T^{2\text{(eff)}}_{2})dr.
\end{equation}
The mass obtained using Tolman's formula coincides with the
Misner-Sharp mass at the boundary but varies within the sphere
(except in the special case of homogeneous distribution and
isotropic pressures). The Misner-Sharp formula has been used
extensively in numerical computations of stellar collapse \cite{25*}
but Tolman mass gives a better estimate in case of anisotropic
fluids. Inserting field Eqs.(\ref{6})-(\ref{8}) in (\ref{18}),
Tolman mass reduces to
\begin{equation}\label{19}
m_T=e^{\frac{\nu-\lambda}{2}}\nu'\frac{r^2}{2}.
\end{equation}
Using the above equation with (\ref{7}), the final expression for
Tolman mass turns out to be
\begin{equation}\label{20}
m_T=e^{\frac{\nu+\lambda}{2}}(m-\frac{r^3}{2}T^{1\text{(eff)}}_{1}).
\end{equation}
The gravitational acceleration is calculated by using 4-acceleration
($a_{\nu}$) as
\begin{equation*}
a=-s^{\mu}a_{\nu}=\frac{e^{-\frac{\lambda}{2}}\nu'}{2},
\end{equation*}
which, in accordance with Eq.(\ref{20}), leads to
\begin{equation*}
a=\frac{e^{-\frac{\nu}{2}}m_T}{r^2}.
\end{equation*}
This shows that Tolman mass can also be treated as the active
gravitational mass. After simplifications \cite{24*}, Tolman mass
can be re-expressed as
\begin{equation}\label{21}
m_T=(m_T)_{\Sigma}(\frac{r}{r_{\Sigma}})^{3}-r^3\int_{r}^{r_{\Sigma}}
e^{\frac{\nu+\lambda}{2}}\left[\frac{1}{r}(T^{1\text{(eff)}}_{1}
-T^{2\text{(eff)}}_{2})+\frac{1}{2r^{4}}\int_{0}^{r}r^{3}
T^{0'\text{(eff)}}_{0}dr\right]dr,
\end{equation}
or equivalently
\begin{equation}\label{22}
m_T=(m_T)_{\Sigma}(\frac{r}{r_{\Sigma}})^{3}-r^3\int_{r}^{r_{\Sigma}}
\frac{e^{\frac{\nu+\lambda}{2}}}{r}\left(\frac{T^{1\text{(eff)}}_{1}
-T^{2\text{(eff)}}_{2}}{2}-\varepsilon\right)dr.
\end{equation}

\section{Structure Scalars}

In order to incorporate the characteristics of the comoving
congruence, we obtain invariants via orthogonal splitting of the
Riemann tensor. These invariants are known as structure scalars. For
this purpose, we use Bel's technique \cite{25} and introduce the
elements of splitting as
\begin{eqnarray*}\label{23}
Y_{\alpha\beta}&=&R_{\alpha\delta\beta\gamma}u^{\delta}u^{\gamma},\\\label{24}
Z_{\alpha\beta}=^*R_{\alpha\delta\beta\gamma}u^{\delta}u^{\gamma}&=&
\frac{1}{2}\eta_{\alpha\delta\mu\epsilon}R^{\mu\epsilon}_{\beta\gamma}
u^{\delta}u^{\gamma},\\\label{25}
X_{\alpha\beta}=^{*}R^{*}_{\alpha\delta\beta\gamma}u^{\delta}u^{\gamma}&=&
\frac{1}{2}\eta_{\alpha\delta}^{\mu\epsilon}R^{*}_{\mu\epsilon\beta\gamma}
u^{\delta}u^{\gamma},
\end{eqnarray*}
where
$R^{*}_{\alpha\beta\delta\gamma}=\frac{1}{2}\eta_{\mu\epsilon\delta\gamma}
R_{\alpha\beta}^{\mu\epsilon}$ and
$^{*}R_{\alpha\beta\delta\gamma}=\frac{1}{2}\eta_{\alpha\beta\mu\epsilon}
R_{\delta\gamma}^{\mu\epsilon}$ are the right and left duals,
respectively. Using the field equations in Eq.(\ref{1'}), the
Riemann tensor takes the form
\begin{equation}\label{26}
R^{\alpha\delta}_{\beta\gamma}=C^{\alpha\delta}_{\beta\gamma}+
2T^{(eff)[\alpha}_{[\beta}\delta^{\delta]}_{\gamma]}+ T^{(eff)}
\left(\frac{1}{3}\delta^{\alpha}_{[\beta}\delta^{\delta}_{\gamma]}-
\delta^{[\alpha}_{[\beta}\delta^{\delta]}_{\gamma]}\right),
\end{equation}
which is split using Eqs.(\ref{3}), (\ref{5}) and (\ref{12}) as
\begin{equation}\label{27}
R^{\alpha\delta}_{\beta\gamma}=R^{\alpha\delta}_{(I)\beta\gamma}
+R^{\alpha\delta}_{(II)\beta\gamma}+R^{\alpha\delta}_{(III)\beta\gamma}
+R^{\alpha\delta}_{(IV)\beta\gamma}+R^{\alpha\delta}_{(V)\beta\gamma},
\end{equation}
where
\begin{eqnarray}\nonumber
R^{\alpha\delta}_{(I)\beta\gamma}&=&\frac{2}{\Phi}\left[\rho
u^{[\alpha}u_{[\beta}\delta_{\gamma]}^{\delta]}
-Ph^{[\alpha}_{[\beta}\delta^{\delta]}_{\gamma]}+(\rho-3P)(\frac{1}{3}
\delta^{\alpha}_{[\beta}\delta^{\delta}_{\gamma]}-
\delta^{[\alpha}_{[\beta}\delta^{\delta]}_{\gamma]})\right],\\\nonumber
R^{\alpha\delta}_{(II)\beta\gamma}&=&\frac{2}{\Phi}
\Pi^{[\alpha}_{[\beta}\delta^{\delta]}_{\gamma]},\\\nonumber
R^{\alpha\delta}_{(III)\beta\gamma}&=&4u^{[\alpha}u_{[\beta}E^{\delta]}_{\gamma]}
-\epsilon^{\alpha\delta}_{\mu}\epsilon_{\beta\gamma\nu}E^{\mu\nu},\\\nonumber
R^{\alpha\delta}_{(IV)\beta\gamma}&=&\frac{2}{\Phi}\left[\Phi^{[,\alpha}_{[;\beta}
\delta^{\delta]}_{\gamma]}+\frac{\omega_{BD}}{\Phi}\Phi^{,[\alpha}\Phi_{,[\beta}
\delta^{\delta]}_{\gamma]}-\left(\Box\Phi+\frac{\omega_{BD}}{2\Phi}\Phi_{,\mu}
\Phi^{,\mu}+\frac{V(\Phi)}{2}\right)\right.\\\nonumber
&\times&\left.\delta^{[\alpha}_{[\beta}\delta^{\delta]}_{\gamma]}\right],
\\\nonumber
R^{\alpha\delta}_{(V)\beta\gamma}&=&-\frac{1}{\Phi}\left[\left(-\frac{\omega_{BD}}
{\Phi}\Phi_{,\mu}\Phi^{,\mu}-2V(\Phi)-3\Box\Phi\right)\left(\frac{1}{3}
\delta^{\alpha}_{[\beta}\delta^{\delta}_{\gamma]}-
\delta^{[\alpha}_{[\beta}\delta^{\delta]}_{\gamma]}\right)\right].
\end{eqnarray}
The three tensors $X_{\alpha\beta},~Y_{\alpha\beta}$ and
$Z_{\alpha\beta}$ are evaluated using the above definition of the
Riemann tensor as
\begin{eqnarray}\nonumber
X_{\alpha\beta}&=&\frac{1}{\Phi}\left(\frac{\rho
h_{\alpha\beta}}{3}+\frac{\Pi_{\alpha\beta}}{2}\right)-E_{\alpha\beta}
-\frac{1}{4\Phi}(\Phi^{,\mu}_{~;\mu}h_{\alpha\beta}-2\Phi_{,\alpha;\mu}u_{\beta}u^{\mu}\\\label{32}
&-&\frac{\omega_{BD}}{{4\Phi^{2}}}\Phi_{,\alpha}\Phi_{,\beta}
+\frac{5h_{\alpha\beta}}{12\Phi}V(\Phi),\\\nonumber
Y_{\alpha\beta}&=&\frac{1}{\Phi}\left(\frac{(\rho+3P)h_{\alpha\beta}}{6}+
\frac{\Pi_{\alpha\beta}}{2}\right)+E_{\alpha\beta}+\frac{1}{2\Phi}(\Phi_{,\alpha;\beta}
-\Phi_{,\alpha;\mu}u_{\beta}u^{\mu}\\\nonumber
&-&\Phi_{,\mu;\beta}u_{\alpha}u^{\mu}
+\Phi_{,\gamma;\mu}u_{\gamma}u^{\mu}g_{\alpha\beta})
+\frac{\omega_{BD}}{2\Phi^2}\Phi_{,\alpha}\Phi_{,\beta}-\frac{h_{\alpha\beta}}{6\Phi}
\left(\frac{\omega_{BD}}{\Phi}\Phi_{,\mu}\Phi^{,\mu}\right.\\\label{33}
&-&\left.V(\Phi)
\right),\\\label{34}Z_{\alpha\beta}&=&\frac{1}{4\Phi}(\eta_{\alpha\gamma\beta\mu}
\Phi^{,\mu}_{~;\delta}u^{\gamma}u^{\delta}).
\end{eqnarray}

Now, the structure scalars \cite{23} are derived by decomposing the
tensors $X_{\alpha\beta}$ and $Y_{\alpha\beta}$ into their trace and
trace-free parts as
\begin{eqnarray*}
X_{\alpha\beta}&=&\frac{X^{\alpha}_{\alpha}}{3}h_{\alpha\beta}+X_{<\alpha\beta>},\\
Y_{\alpha\beta}&=&\frac{Y^{\alpha}_{\alpha}}{3}h_{\alpha\beta}+Y_{<\alpha\beta>},
\end{eqnarray*}
where
\begin{eqnarray*}
X_{<\alpha\beta>}&=&h^{\nu}_{\alpha}h^{\mu}_{\beta}\left(X_{\mu\nu}
-\frac{X^{\alpha}_{\alpha}}{3}h_{\mu\nu}\right),\\
Y_{<\alpha\beta>}&=&h^{\nu}_{\alpha}h^{\mu}_{\beta}\left(Y_{\mu\nu}
-\frac{Y^{\alpha}_{\alpha}}{3}h_{\mu\nu}\right).
\end{eqnarray*}
Denoting $X^{\alpha}_{\alpha}$ by $X_T$ and  $Y^{\alpha}_{\alpha}$
by $Y_T$, the four structure scalars in the presence of scalar field
turn out to be
\begin{eqnarray}\nonumber
X_{T}&=&X_{T}^{m}+X_{T}^{\Phi}=\frac{1}{4\Phi}(\rho)-\frac{1}{4\Phi}\left(5\Box\Phi
-2\Phi_{,\gamma;\mu}u^{\gamma}u^{\mu}
-\frac{\omega_{BD}}{\Phi}\Phi_{,\alpha}\Phi^{,\alpha}\right.\\\label{35}
&+&\left.5V(\Phi)\right),\\\nonumber
X_{TF}&=&X_{TF}^{m}+X_{TF}^{\Phi}=\frac{1}{\Phi}(\frac{\Pi}{2}-\varepsilon\Phi)+\frac{1}{2\Phi}
\left(\Box\Phi-\frac{\omega_{BD}}
{\Phi}\Phi_{,\alpha}\Phi^{,\alpha}\right.\\\label{36}
&-&\left.\Phi_{,\gamma;\mu}u^{\gamma}u^{\mu}\right),\\\nonumber
Y_{T}&=&Y_{T}^{m}+Y_{T}^{\Phi}=\frac{1}{2\Phi}(\rho+3P_r-2\Pi)+\frac{1}{2\Phi}\left(\Box\Phi
+2\Phi_{,\gamma;\mu}u^{\gamma}u^{\mu}\right.\\\label{37}
&+&\left.V(\Phi)\right),\\\nonumber
Y_{TF}&=&Y_{TF}^{m}+Y_{TF}^{\Phi}=\frac{1}{\Phi}(\frac{\Pi}{2}+\varepsilon\Phi)
+\frac{1}{2\Phi} \left(\Box\Phi+\frac{\omega_{BD}}
{\Phi}\Phi_{,\alpha}\Phi^{,\alpha}\right.\\\label{38}
&-&\left.\Phi_{,\gamma;\mu}u^{\gamma}u^{\mu}\right).
\end{eqnarray}
It follows from the above equations that under the influence of
scalar field, the total energy density within the system is
determined by $X_T$ whereas the scalar $Y_T$ describes the effects
of principal stresses produced by inhomogeneous energy density.
Using Eqs.(\ref{36}) and (\ref{38}), we have
\begin{equation*}
X_{TF}^{m}+Y_{TF}^{m}=\frac{2\Pi}{\Phi}~\text{and}~
Y_{TF}^{\Phi}-X_{TF}^{\Phi}=\frac{\omega_{BD}}
{\Phi^2}\Phi_{,\alpha}\Phi^{,\alpha},
\end{equation*}
which shows that local anisotropy in pressure is found by
$X_{TF}^{m}$ and $Y_{TF}^{m}$ whereas the coupling parameter is
determined by $X_{TF}^{\Phi}$ and $Y_{TF}^{\Phi}$. From
Eqs.(\ref{22}) and (\ref{38}), it is observed that $Y_{TF}$ appears
in the expression for Tolman mass as
\begin{equation}\label{39}
m_{T}=(m_{T})_{\Sigma}\left(\frac{r}{r_{\Sigma}}\right)+r^3\int_{r}^{r_{\Sigma}}
\frac{e^{\frac{\nu+\lambda}{2}}}{r}(-Y_{TF}^m+Y_{TF}^\Phi)+
\frac{e^{\frac{\nu-\lambda}{2}}\Phi'}{2r\Phi}dr.
\end{equation}
This indicates that $Y_{TF}$ gauges the impact of anisotropic
pressure and inhomogeneous density on the active gravitational mass.

\section{Complexity Factor}

In this section, we formulate the complexity factor which is
governed by the physical features such as energy density, pressure,
heat flux. In general, a system is said to be least complex if its
physical structure is completely described by a small number of
factors. For example, a spherical object filled with dust fluid has
only one necessary ingredient which is the energy density of the
fluid whereas, the inclusion of isotropic pressure to dust fluid
leads to a slightly more complex system known as a perfect fluid. In
GR, the complexity factor depends on the inhomogeneous and
anisotropic distribution \cite{13}. However, in our work, the
complexity is determined by the scalar field and self-interacting
scalar potential in addition to inhomogeneous energy density and
anisotropic pressure. The complexity of this system can, therefore,
be completely described by the structure scalar $Y_{TF}$, since it
is not only a relation between the sources of complexity but also a
measure of how they affect the Tolman mass. Setting $Y_{TF}=0$ leads
to vanishing complexity factor condition which establishes the
following relation among the physical variables
\begin{equation}\label{40}
\frac{\Pi}{\Phi}=\frac{1}{2r^3}\int_{0}^{r}r^{3}T^{0'(eff)}_{0}dr+
\frac{e^{-\lambda}\Phi'}{2r\Phi}.
\end{equation}
This condition can be used as a restraint for formulating solution
of the field equations.

Gokhroo and Mehra \cite{26} obtained a physically reasonable
interior solution for an anisotropic sphere with variable energy
density to explain the larger red-shifts of quasi-stellar objects.
Using their assumptions, we illustrate the behavior of
self-gravitating system for the condition of vanishing complexity.
The assumed energy density (maximum at the center and decreasing
along the radius) is given by
\begin{equation}\label{42}
\rho=\rho_{0}(1-\frac{r^2}{r^2_{\Sigma}}),
\end{equation}
which leads to the mass function
\begin{equation}\label{43}
m(r)=\frac{1}{2}\left[\frac{\rho_{0}r^3}{3\Phi}\left(1-\frac{3kr^2}{5r^2_{\Sigma}}\right)
+\int_{0}^{r}\frac{r^2}{\Phi}T_{0}^{0\Phi}dr\right].
\end{equation}
Substituting the above equation in (\ref{9}), the expression for the
metric function turns out to be
\begin{equation}\label{41}
e^{-\lambda}=\frac{1}{\Phi}\left(1-\beta r^2+\frac{3k\beta
r^4}{5r^2_{\Sigma}}\right)
-\int_{0}^{r}\frac{r^2}{\Phi}T_{0}^{0\Phi}dr,
\end{equation}
where $k\in(0,1)$ and $\beta=\frac{\rho_{0}}{3}$. Using
Eqs.(\ref{7}) and (\ref{8}), it follows that
\begin{eqnarray}\nonumber
&&\frac{1}{\Phi}\{\Pi+e^{-\lambda}[\Phi''+\Phi'(-\frac{\lambda'}{2}
+\frac{1}{r})+\frac{\omega_{BD}}{\Phi}\Phi'^2]\}=e^{-\lambda}[\frac{-\nu''}{2}
-\frac{\nu'^{2}}{4}\\\label{44}
&&+\frac{\nu'}{2r}+\frac{1}{r^2}+\frac{\lambda'}{2}
(\frac{\nu'}{2}+\frac{1}{r})]-\frac{1}{r^2}.
\end{eqnarray}
Introducing new variables as
\begin{eqnarray}\label{45}
e^{\nu}&=&e^{\int(2z-\frac{2}{r})dr},\\\label{46}
e^{-\lambda}=y(r)&=&\frac{1}{\Phi}\left(1-\beta r^2+\frac{3k\beta
r^4}{5r^2_{\Sigma}}\right)
-\int_{0}^{r}\frac{r^2}{\Phi}T_{0}^{0\Phi}dr,
\end{eqnarray}
such that Eq.(\ref{44}) reduces to
\begin{eqnarray}\nonumber
\left(-\frac{2\Phi}{\Phi
z+\Phi'}\right)\left(\frac{\Pi}{2\Phi}+\frac{1}{r^2}\right)&=&y'
+y\left(\frac{2\Phi}{\Phi
z+\Phi'}\right)\left[z^2-\frac{3z}{r}+z'+\frac{2}{r^2}\right.\\\label{47}
&+&\left.\frac{1}{\Phi}\left(\Phi''+\frac{\Phi'}{r}
+\frac{\omega_{BD}}{\Phi}\Phi'^{2}\right)\right],
\end{eqnarray}
with the value of $\Pi$ provided by Eqs.(\ref{40}) and (\ref{42}).
Hence, the metric can be expressed in terms of the new variables as
\begin{equation}\nonumber
ds^2=-e^{\int(2z-\frac{2}{r})dr}+\frac{\xi}{\int{\left(-\frac{2\Phi}{\Phi
z+\Phi'}\right)}\left(\frac{\Pi}{2\Phi}+\frac{1}{r^2}\right)\xi
dr+C}dr^2+r^2d\theta^2+r^2\sin^2\theta d\phi^2,
\end{equation}
where $C$ is a constant of integration and
\begin{eqnarray*}
\xi&=&\text{exp}\left\{\int\left(\frac{2\Phi}{\Phi
z+\Phi'}\right)\left[z^2-\frac{3z}{r}+z'+\frac{2}{r^2}\right.\right.\\
&+&\left.\left.\frac{1}{\Phi}\left(\Phi''+\frac{\Phi'}{r}
+\frac{\omega_{BD}}{\Phi}\Phi'^{2}\right)\right]dr\right\}.
\end{eqnarray*}
The energy density, radial and tangential pressure in the presence
of scalar field take the form
\begin{eqnarray}\nonumber
\rho&=&\frac{2\Phi
m'}{r^2}-(1-\frac{2m}{r})\left[\left(\frac{2}{r}+\frac{m'}{(r-2m)}\right)\Phi'
+\Phi''-\frac{\omega_{BD}}{2\Phi}\Phi'^2\right]\\\nonumber
&+&\frac{V(\Phi)}{2},\\\nonumber
P_r&=&\frac{\Phi}{2r^2}\left[-1+\frac{m}{r}+z(r-2m)\right]-(1-\frac{2m}{r})
\left[\left(-\frac{1}{r}-z\right)\Phi'+\Phi''\right.\\\nonumber
&+&\left.\frac{\omega_{BD}}{2\Phi}\Phi'^2\right]-\frac{V(\Phi)}{2},\\\nonumber
\text{and}\\\nonumber
P_{\perp}&=&\frac{\Phi}{2}\left[(\frac{1}{r^2}+z^2+z'-\frac{z}{r})
+\frac{z}{2}(\frac{m}{r^2}-\frac{m'}{r})\right]-(1-\frac{2m}{r})\\\nonumber
&\times&\left[\left(z-\frac{m'}{(r-2m)}\right)\Phi'
-\Phi''-\frac{\omega_{BD}}{2\Phi}\Phi'^2\right]-\frac{V(\Phi)}{2}.
\end{eqnarray}

The interior of self-gravitating systems depends on a large number
of state variables such as pressure, energy density, temperature,
etc. However, not all factors need to have the same impact on the
internal structure. The dominant factors in a given scenario are
often related through an equation of state (EoS) which helps in the
study of compact objects. The polytropic EoS with anisotropic
effects have been used extensively in the study of stellar objects
\cite{27} which is defined as
\begin{equation}\label{47*}
P_r=K\rho^{\gamma}=K\rho^{\frac{n+1}{n}},
\end{equation}
where $K,~n$ and $\gamma$ denote polytropic constant, polytropic
index and polytropic exponent, respectively. Introducing new
variables
\begin{equation*}
\alpha=\frac{P_{r0}}{\rho_0},\quad r=\frac{\xi}{A},\quad
A^2=\frac{\rho_0}{2\alpha(n+1)},\quad
\psi^n=\frac{\rho}{\rho_0},\quad v(\xi)=\frac{2m(r)A^3}{\rho_0},
\end{equation*}
lead to the following form of Eqs.(\ref{2*})-(\ref{8*})
\begin{eqnarray}\nonumber
\Phi&=&\frac{-\left[1-\frac{2(n+1)\alpha
v}{\xi}\right]^{-1}}{(3+2\omega_{BD})}\left\{\int
\frac{\rho_0}{(2A\xi)^2}\left[v+\frac{\alpha\xi^3\psi^{n+1}}{2\Phi}
-\frac{\xi^3T_{1}^{\Phi1}}{2\rho_0\Phi}\right]\right.\\\nonumber
&-&\left.\frac{\rho_0v}{2(A\xi)^2}\left[1-\frac{2(n+1)\alpha
v}{\xi}\right]^{-1}d\xi
\left[\rho_0\psi^n(1-3k\psi\rho_0^{\frac{1}{n}})
+\Phi\frac{dV(\Phi)}{d\Phi}\right.\right.\\\label{49*}
&-&\left.\left.2V(\Phi)\right]\right\},\\\label{49}
\frac{dv}{d\xi}&=&\frac{\xi^2\psi^{n}}{\Phi}+\frac{\xi^2}{\Phi\rho_{0}}T_0^{\Phi0},\\\nonumber
\frac{d\psi}{d\xi}
&=&\frac{\psi^{-n}}{\alpha(n+1)}\frac{dT_1^{\Phi1}}{d\xi}
-\left\{\frac{\xi^2}{2}\left[\frac{1-\frac{2(n+1)\alpha
v}{\xi}}{1+\alpha\psi}\right]\right\}^{-1}
\left[\left(v+\frac{\alpha\xi^3\psi^{n+1}}{2\Phi}\right.\right.\\\nonumber
&-&\left.\left.\frac{\xi^3}{2\rho_0\Phi}T_1^{\Phi1}\right)
\left(1+\frac{T_0^{\Phi0}-T_1^{\Phi1}}{\Phi}\right)
+\frac{\xi}{n+1}\left(\frac{1-\frac{2(n+1)\alpha
v}{\xi}}{1-\alpha\psi}\right)\right.\\\label{50}
&\times&\left.\left(\frac{\psi^{-n}}{P_{r0}}\Pi
+\frac{\Upsilon}{\alpha}\right)\right],
\end{eqnarray}
where
\begin{eqnarray*}
T_0^{\Phi0}&=&\left[1-\frac{2v\alpha(n+1)}{\xi}\right]
\left[\Phi''+\frac{\omega_{BD}\Phi'^2}{2\Phi}
+\frac{2A\Phi'}{\xi}\right]+\frac{\rho_0v\Phi'}{2A\xi^2}+\frac{V(\Phi)}{2},\\
T_1^{\Phi1}&=&\left(\frac{4}{1+r\Phi'}\right)\left\{\frac{\Phi'}{4A\xi^2}
\left(\rho_0v+\xi^3K\psi^{n+1}\rho_0^{\gamma}\right)+
\left[1-2\frac{v\alpha(n+1)}{\xi}\right]\right.\\
&\times&\left.\left(\frac{2A\Phi'}{\xi}
-\frac{\omega_{BD}\Phi'^2}{2\Phi}\right)-\frac{V(\Phi)}{2}\right\},\\
\Upsilon&=&\left[1-\frac{2v\alpha(n+1)}{\xi}\right]
\left[\Phi''+\frac{\omega_{BD}\Phi'^2}{2\Phi}
-\frac{A\Phi'}{\xi}\right]+\frac{\rho_0v\Phi'}{2A\xi^2}.
\end{eqnarray*}
The subscript $0$ indicates the behavior of respective quantities at
the center. The vanishing complexity condition in terms of the new
variables is
\begin{eqnarray}
\frac{6\Pi}{n\rho_0}+\frac{2\xi}{n\rho_0}\frac{d\Pi}{d\xi}
&=&\xi\psi^{n-1}\frac{d\psi}{d\xi}
+\frac{\xi}{n\rho_0}\frac{dT_0^{\Phi0}}{d\xi}+\frac{\rho_0vA^2}{2\Phi\xi}\frac{d\Phi}{d\xi}
+\frac{\xi A^2}{\Phi}\\\label{51}
&\times&\left[1-\frac{2v\alpha(n+1)}{\xi}\right]
\left(\frac{d\Phi}{d\xi} +\frac{A}{2}\frac{d^2\Phi}{d\xi^2}\right),
\end{eqnarray}
where
\begin{eqnarray*}
\frac{dT_0^{\Phi0}}{d\xi}&=&
\left[\frac{2v\alpha(n+1)}{\xi^2}\right]
\left(A^2\frac{d^2\Phi}{d\xi^2}+\frac{A\omega_{BD}}{\Phi}\frac{d\Phi}{d\xi}
+\frac{2A^2}{\xi}\frac{d\Phi}{d\xi}\right)\\
&+&\left[1-\frac{2v\alpha(n+1)}{\xi}\right]
\left[A^3\frac{d^3\Phi}{d\xi^3}+\frac{2\omega_{BD}A^2}{\Phi}\frac{d\Phi}{d\xi}
{d^2\Phi}{d\xi^2}-\frac{\omega_{BD}A^2}{\Phi^2}\right.\\
&\times&\left.\left(\frac{d\Phi}{d\xi}\right)^2
-\frac{2A^2}{\xi^2}\frac{d\Phi}{d\xi}+2A^3\frac{d^2\Phi}{d\xi^2}\right]+\frac{\rho_0v}{2\xi^3}
\left(\xi\frac{d^2\Phi}{d\xi^2}-\frac{d\Phi}{d\xi}\right)\\
&+&\frac{\xi}{2n\rho_0}\frac{dV(\Phi)}{d\Phi}\frac{d\Phi}{d\xi}.
\end{eqnarray*}
The system of four equations (\ref{49*})-(\ref{51}) in five unknowns
($\Pi,~v,~\psi,~\Phi,~ V(\Phi)$) gives us the freedom to fix one of
the unknowns. Hence, the solutions of the system will vary according
to the choice of $V(\Phi)$. Further, the obtained solutions can be
checked for viability and stability through energy conditions
\cite{18**},\cite{29} and Chandrasekhar technique \cite{30},
respectively.

\section{Summary}

In the field of astrophysics, the class of self-gravitating fluids
is characterized by long-range gravitational interactions that have
tempted researchers to explore their physical properties
(luminosity, mass-radius ratio, stability, etc.). Studies have shown
that such systems exhibit interesting but complex dynamical
behavior. This work is devoted to formulating the complexity factor
of a static sphere filled with anisotropic fluid in the framework of
self-interacting BD theory. We have used Misner-Sharp and Tolman
definitions for calculating the mass of the sphere. We measure the
complexity of the self-gravitating objects through a scalar function
which incorporates the effects of comoving congruence. For this
purpose, the orthogonal splitting of the Riemann tensor is used to
develop four structure scalars, $X_T,~X_{TF},~Y_T$, and $Y_{TF}$.
Moreover, a relation between anisotropic pressure and inhomogeneous
energy density has been evaluated through the vanishing complexity
constraint. We have used this condition to check the behavior of a
variable energy density model developed by Gokhroo and Mehra
\cite{26}. Finally, the polytropic EoS is implemented along with the
vanishing complexity condition to find a possible solution for the
system.

Self-interacting BD theory modifies Einstein-Hilbert action by
allowing a dynamical gravitational constant in terms of a scalar
field $\Phi$. The effective potential function $V(\Phi)$ adjusts the
values of cosmic inflation with the observational data. The
structure scalars contain additional effects of the scalar field and
potential function. Consequently, a complexity factor in terms of
these scalars includes scalar field as a source of complexity. Here,
the simplest configuration corresponds to two cases of fluids
(either the fluid is homogeneous and isotropic or the inhomogeneity
and anisotropy terms cancel each other) which must also satisfy an
additional condition ($\frac{e^{-\lambda}\Phi'}{2r\Phi}=0$). It has
also been observed that the constraint of vanishing complexity
condition reduces the number of degrees of freedom while obtaining a
solution to the self-interacting BD field equations.

It is found that physical variables (inhomogeneous energy density
and anisotropic pressure), scalar field and self-interacting
potential determine the complexity of the fluid. Among the proposed
structure scalars, $Y_{TF}$ is the most appropriate complexity
factor as it establishes a relation between these factors as well as
measures their effect on Tolman mass. In the case of homogeneous
energy density and isotropic pressure, the complexity factor of the
system vanishes in the context of GR \cite{13}. However, our results
do not deduce to a complexity free structure under these assumptions
which highlight the effect of including scalar field and potential
function to the complexity factor. Hence, the presence of scalar
field and self-interacting potential has increased the complexity of
the system. The use of vanishing complexity condition in two models
representing self-gravitating objects have provided open systems
which can be closed by assigning suitable values to $V(\Phi)$
\cite{28}. It is interesting to mention here that all the results
presented in this paper reduce to GR \cite{13} under the conditions
$\Phi=\text{constant}$ and $\omega_{BD}\rightarrow \infty$.

\end{document}